\documentstyle[12pt]{article}

\makeatletter
\def\@cite#1#2{$^{\hbox{\scriptsize{#1\if@tempswa , #2\fi}}}$}
\makeatother

\newlength{\minitwocolumn}
\catcode`\@=11
\long\def\@makefntext#1{
\protect\noindent \hbox to 3.2pt {\hskip-.9pt  
$^{{\eightrm\@thefnmark}}$\hfil}#1\hfill}		

\def\@makefnmark{\hbox to 0pt{$^{\@thefnmark}$\hss}}	
	
\def\ps@myheadings{\let\@mkboth\@gobbletwo
\def\@oddhead{\hbox{}
\rightmark\hfil\eightrm\thepage}   
\def\@oddfoot{}\def\@evenhead{\eightrm\thepage\hfil
\leftmark\hbox{}}\def\@evenfoot{}
\def\sectionmark##1{}\def\subsectionmark##1{}}

\font\eightrm=cmr8

\mathchardef\alpha="710B
\mathchardef\beta="710C
\mathchardef\gamma="710D
\mathchardef\delta="710E
\mathchardef\epsilon="710F
\mathchardef\zeta="7110
\mathchardef\eta="7111
\mathchardef\theta="7112
\mathchardef\iota="7113
\mathchardef\kappa="7114
\mathchardef\lambda="7115
\mathchardef\mu="7116
\mathchardef\nu="7117
\mathchardef\xi="7118
\mathchardef\pi="7119
\mathchardef\rho="711A
\mathchardef\sigma="711B
\mathchardef\tau="711C
\mathchardef\upsilon="711D
\mathchardef\phi="711E
\mathchardef\chi="711F
\mathchardef\psi="7120
\mathchardef\omega="7121
\mathchardef\varepsilon="7122
\mathchardef\vartheta="7123
\mathchardef\varpi="7124
\mathchardef\varrho="7125
\mathchardef\varsigma="7126
\mathchardef\varphi="7127

\font\fivmib = cmmib10  \@ptscale5 
\font\sixmib = cmmib10  \@ptscale6 
\font\sevmib = cmmib10  \@ptscale7 
\font\egtmib = cmmib10  \@ptscale8 
\font\ninmib = cmmib10  \@ptscale9 
\font\tenmib = cmmib10   
\font\elvmib = cmmib10   \@halfmag 
\font\twlmib = cmmib10   \@magscale1 
\font\frtnmib = cmmib10   \@magscale2 
\font\svtnmib = cmmib10   \@magscale3 
\font\twtymib = cmmib10   \@magscale4 
\font\twfvmib = cmmib10   \@magscale5 

\def\mbf{%
\mathcode`0="0630
\mathcode`1="0631
\mathcode`2="0632
\mathcode`3="0633
\mathcode`4="0634
\mathcode`5="0635
\mathcode`6="0636
\mathcode`7="0637
\mathcode`8="0638
\mathcode`9="0639
\protect\pmib}
\newfam\mibfam
\newcount\mibf@mnum
\mibf@mnum=\mibfam \multiply\mibf@mnum by 256 
\def\mbfplus{%
\mathcode`+="262B
\mathcode`-="272D
\mathcode`=="363D
\mathcode`/="713D
\mathcode`(="4628
\mathcode`[="465B
\mathcode`)="5629
\mathcode`]="565D
\advance\mibf@mnum by "303C 
\mathcode`<=\mibf@mnum
\advance\mibf@mnum by 2
\mathcode`>=\mibf@mnum
}

\newif\ifMB@full \MB@fulltrue
\def\subfontMBF{\MB@fullfalse}
\let\MB@vpt\vpt
\let\MB@vipt\vipt
\let\MB@viipt\viipt
\let\MB@viiipt\viiipt
\let\MB@ixpt\ixpt
\let\MB@xpt\xpt
\let\MB@xipt\xipt
\let\MB@xiipt\xiipt
\let\MB@xivpt\xivpt
\let\MB@xviipt\xviipt
\let\MB@xxpt\xxpt
\let\MB@xxvpt\xxvpt

\def\vpt{\MB@vpt \def\boldmath{}%
\ifMB@full
\def\pmib{\fam\mibfam\fivmib}\textfont\mibfam\fivmib 
   \scriptfont\mibfam\fivmib \scriptscriptfont\mibfam\fivmib 
\else \def\pmib{} \fi}

\def\vipt{\MB@vipt \def\boldmath{}%
\ifMB@full
\def\pmib{\fam\mibfam\sixmib}\textfont\mibfam\sixmib 
   \scriptfont\mibfam\fivmib \scriptscriptfont\mibfam\fivmib 
\else \def\pmib{} \fi}

\def\viipt{\MB@viipt \def\boldmath{}%
\ifMB@full
\def\pmib{\fam\mibfam\sevmib}\textfont\mibfam\sevmib 
   \scriptfont\mibfam\sixmib \scriptscriptfont\mibfam\fivmib 
\else \def\pmib{} \fi}

\def\viiipt{\MB@viiipt \def\boldmath{}%
\ifMB@full
\def\pmib{\fam\mibfam\egtmib}\textfont\mibfam\egtmib 
   \scriptfont\mibfam\sixmib \scriptscriptfont\mibfam\fivmib 
\else \def\pmib{} \fi}

\def\ixpt{\MB@ixpt \def\boldmath{}%
\ifMB@full
\def\pmib{\fam\mibfam\ninmib}\textfont\mibfam\ninmib 
   \scriptfont\mibfam\sixmib \scriptscriptfont\mibfam\fivmib 
\else \def\pmib{} \fi}

\def\xpt{\MB@xpt \def\boldmath{}%
\ifMB@full
\def\pmib{\fam\mibfam\tenmib}\textfont\mibfam\tenmib 
    \scriptfont\mibfam\sevmib \scriptscriptfont\mibfam\fivmib 
\else
\def\pmib{\fam\mibfam\tenmib}\textfont\mibfam\twlmib 
    \scriptfont\mibfam\tenmib \scriptscriptfont\mibfam\tenmib 
\fi}

\def\xipt{\MB@xipt \def\boldmath{}%
\ifMB@full
\def\pmib{\fam\mibfam\elvmib}\textfont\mibfam\elvmib 
   \scriptfont\mibfam\egtmib \scriptscriptfont\mibfam\sixmib 
\else
\def\pmib{\fam\mibfam\elvmib}\textfont\mibfam\twlmib 
    \scriptfont\mibfam\tenmib \scriptscriptfont\mibfam\tenmib 
\fi}

\def\xiipt{\MB@xiipt \def\boldmath{}%
\ifMB@full
\def\pmib{\fam\mibfam\twlmib}\textfont\mibfam\twlmib
   \scriptfont\mibfam\egtmib \scriptscriptfont\mibfam\sixmib
\else
\def\pmib{\fam\mibfam\twlmib}\textfont\mibfam\twlmib 
    \scriptfont\mibfam\tenmib \scriptscriptfont\mibfam\tenmib 
\fi}

\def\xivpt{\MB@xivpt \def\boldmath{}%
\ifMB@full
\def\pmib{\fam\mibfam\frtnmib}\textfont\mibfam\frtnmib
   \scriptfont\mibfam\tenmib \scriptscriptfont\mibfam\sevmib
\else
\def\pmib{\fam\mibfam\frtnmib}\textfont\mibfam\frtnmib
    \scriptfont\mibfam\elvmib \scriptscriptfont\mibfam\tenmib 
\fi}

\def\xviipt{\MB@xviipt \def\boldmath{}%
\ifMB@full
\def\pmib{\fam\mibfam\svtnmib}\textfont\mibfam\svtnmib
    \scriptfont\mibfam\twlmib \scriptscriptfont\mibfam\tenmib 
\else \def\pmib{} \fi}

\def\xxpt{\MB@xxpt \def\boldmath{}%
\ifMB@full
\def\pmib{\fam\mibfam\twtymib}\textfont\mibfam\twtymib
    \scriptfont\mibfam\frtnmib \scriptscriptfont\mibfam\twlmib 
\else \def\pmib{} \fi}

\def\xxvpt{\MB@xxvpt \def\boldmath{}%
\ifMB@full
\def\pmib{\fam\mibfam\twfvmib}\textfont\mibfam\twfvmib
    \scriptfont\mibfam\twtymib \scriptscriptfont\mibfam\svtnmib 
\else \def\pmib{} \fi}

\font\sc=cmr5 scaled\magstep1

\begin{document}



\pagestyle{empty}

\begin{center}
{\large\bf 
How to Solve Quantum Nonlinear Abelian Gauge Theory\\
in Two Dimension in the Heisenberg Picture\\
\vskip 1mm
}

\vspace{15mm}

Noriaki IKEDA
\footnote{ E-mail address:\ nori@kurims.kyoto-u.ac.jp } \\
Research Institute for Mathematical Sciences \\
Kyoto University, Kyoto 606-01, Japan
\end{center}
\date{}


\vspace{15mm}
\begin{abstract}
The new method based on the operator formalism proposed by
Abe and Nakanishi is applied to the quantum nonlinear abelian gauge
theory in two dimension.
The soluble models in this method are extended to wider class of quantum
field theories.
We obtain the exact solution
in the canonical-quantization
operator formalism in the Heisenberg picture.
So this analysis might shed some light on
the analysis of gravitational theory and
non-polynomial field theories.
\end{abstract}

\newpage
\pagestyle{plain}
\pagenumbering{arabic}


\rm
\section{Introduction}
\noindent
Quantum Einstein gravity is non-renormalizable in the conventional
perturbative approach.
One candidate to overcome its difficulty is to modify the perturbative 
approach.
In [1], Abe and Nakanishi
have proposed the new method to solve quantum field theory in the
operator formalism in the Heisenberg picture.
The procedure is the following. 

First we calculate equal-time commutation relations of the fundamental
fields from the canonical commutation relations. 
~From equal-time commutation relations and the equations of motion,
we set up the Cauchy problems for two-dimensional commutation relations.
By solving the Cauchy problems with operator coefficients,
we obtain the two-dimensional commutation relations of fundamental
fields.
Finally, we construct full Wightman functions from the consistency
with multiple commutation
relations for fundamental fields and from the energy positivity
requirement.
This enables us to separate algebraic relations of operators
and its realization or reguralization on a state space.

So far, in the theories which have been solved exactly by this method
there is at least one fundamental field, $\phi(x)$, which commutes
mutually: 
\begin{eqnarray}
[ \phi(x), \phi(y)] = 0,
\label{phicom}
\end{eqnarray}
for arbitrary $x$ and $y$. 
This full commutativity is crucial in solvability of such models as 
two-dimensional quantum gravity in covariant gauge or light-cone gauge,
BF theory and the one-loop model\cite{AbeNak,AbeNak2}.
In this paper, we extend the above method to a model with more 
general commutation relations, that is, the model in which there is no 
such a fundamental fields as $\phi(x)$ satisfying (\ref{phicom}).


The non-polynomiality is a common feature of the gravitational theories,
which makes their analysis difficult. 
Moreover they are generally non-renormalizable.
So it is unnatural to analyze it in the conventional 
perturbative approach based on the interaction picture.
A natural method is to treat it in the Heisenberg picture.

In this paper, we consider a general $U(1)$ gauge theory in two dimension
as a toy model 
which can be solved in the Heisenberg operator formalism.
Maxwell's theory can be generalized to 
the nonlinear electromagnetic theory of a massless vector $A_{\mu}$
with $U(1)$ gauge symmetry as 
\begin{eqnarray}
S & = & \int d^2 x {\cal L}, \nonumber \\
{\cal L} & = & \Phi(F_{\mu\nu} F^{\mu\nu}) + \mbox{matters},
\end{eqnarray}
where $\Phi(z)$ is a function of $z$.
The Born-Infeld theory\cite{BI} is one example of 
nonlinear electromagnetic theory.



This Lagrangian density 
is generally non-polynomial in the 
fundamental fields, 
Thus the analysis of this theory might also give us a key to
investigate the above problems.

In this paper, we exactly solve two-dimensional
quantum nonlinear abelian gauge theory 
in the light-cone gauge.

We calculate all the exact multiple commutation relations and 
all n-point Wightman functions for the electromagnetic field.

\vskip50pt

\def\vev#1{{\langle#1\rangle}}
\def\calL{{\cal L}}
\def\brs{{\delta_B}}
\def\aplus{{A_{+}}}
\def\aminus{{A_{-}}}
\def\Dplus{{D^{(+)}}}
\def\Dplusin{{D^{(+)}_{<}}}
\def\wight#1{\langle#1\rangle}
\def\trun#1{\langle#1\rangle_{\hbox{\sc T}}}
\def\Da{{D}}
\def\Dplusa{{D^{(+)}}}
\def\Dplusina{{D^{(+)}_{<}}}

\def\fvev{{\mbf f}}
\def\gvev{{\mbf g}}
\def\sym{{\cal S}}

\section{Nonlinear Abelian Gauge Theory}
\noindent
The action of the nonlinear abelian gauge theory in two dimension
is written as 
\begin{eqnarray}
S & = & \int d^2 x {\cal L}, \nonumber \\ 
{\cal L} & = & \Phi(F_{\mu\nu} F^{\mu\nu}),
\label{Lag}
\end{eqnarray}
where $\Phi(z)$ is a function of $z$.
The property of $\Phi(z)$ is specified later.
If $\Phi(z) = - \frac{1}{4} z$, 
we obtain the Maxwell theory.
%
\begin{eqnarray}
{\cal L} = - \frac{1}{4} F_{\mu\nu} F^{\mu\nu}.
\end{eqnarray}
However the Lagrangian density generally becomes the infinite series
of the fundamental fields.

In order to carry out canonical quantization, we fix the 
gauge of $U(1)$ symmetry. We take the light-cone gauge:
$\aminus = 0$, where 
\begin{eqnarray}
A_{\pm} = A_0 \pm A_1.
\end{eqnarray}
Then (\ref{Lag}) is written as 
%
%
%
%
%
\begin{eqnarray}
{\cal L} & = & \Phi ( - 2
(\partial_{-} A_{+} )^2).
\label{Lagfix}
\end{eqnarray}
where $x^\pm = x^0 \pm x^1$.
The equations of motion are derived from (\ref{Lagfix}) as follows:
\begin{eqnarray}
&& \partial_{-} \left(
\frac{\partial \Phi}{\partial (\partial_{-} \aplus )}
\right) = 0, 
\label{eqA}
\end{eqnarray}
The canonical conjugate momentum of $A_{+}$ is
\begin{eqnarray}
\pi_{\aplus} \equiv 
\frac{\partial {\cal L}}{\partial (\partial_0 A_+)}
= \frac{1}{2} \frac{\partial \Phi}
{\partial (\partial_{-} \aplus )}.
\label{pieq}
\end{eqnarray}
If we define
\begin{eqnarray}
\Psi(\partial_{-} \aplus) \equiv \frac{\partial \Phi}{\partial
(\partial_{-} \aplus )},
\end{eqnarray}
then (\ref{pieq}) is rewritten as
\begin{eqnarray}
\pi_{\aplus} = \frac{1}{2} \Psi(\partial_{-} \aplus).
\label{pia}
\end{eqnarray}

Let us analyze the solution of this theory.
In order to quantize the theory, we set up the canonical commutation
relations of the canonical quantities as follows:
\begin{eqnarray}
&& [ \pi_{\aplus}, \aplus ]|_{0} = - i \delta( x^1 - y^1), \nonumber \\
&& [ \aplus, \aplus ]|_{0}  = 0, \nonumber \\
&& [ \pi_{\aplus}, \pi_{\aplus} ]|_{0} = 0,
\label{cancom}
\end{eqnarray}
where $[\ ,\ ]|_{0} $ denotes the equal-time commutation relations at 
$x^0 = y^0$. 
We want to calculate the multiple commutation relations 
\begin{eqnarray}
[ \cdots[ [ \aplus(x_1), \aplus(x_2) ], \aplus(x_3) ], 
\cdots, \aplus(x_n)].
\label{mcom}
\end{eqnarray}
In the precise Abe-Nakanishi method,
we set up the Cauchy problems of (\ref{mcom}) from (\ref{eqA}),
(\ref{pia}) and (\ref{cancom}) and solve them directly.
We can obtain the multiple commutation relations in the above method,
but we use simpler method in this paper.

We can solve the equations of motion explicitly. 
Integrating (\ref{eqA}), we can write
\begin{eqnarray}
\frac{\partial \Phi}{\partial (\partial_{-} \aplus )}
= \Psi(\partial_{-} \aplus ) = f(x^+),
\label{1int}
\end{eqnarray}
where $f(x^+)$ is a 
hermitian operator depending on only $x^+$.
Here we assume $\Psi$ be invertible.
Then (\ref{1int}) is rewritten as
\begin{eqnarray}
\partial_{-} \aplus(x) = \Psi^{-1}(f(x^+)).
\label{Arepf}
\end{eqnarray}
Therefore we can solve $\aplus(x)$ as
\begin{eqnarray}
\aplus(x) = \Psi^{-1}(f(x^+)) x^- + g(x^+),
\label{aplusrep}
\end{eqnarray}
where $g(x^+)$ is a hermitian operator depending only on $x^+$.
%
%

We can express 
$\pi_{\aplus}$ in terms of  $f(x^+)$ from (\ref{pia}) and (\ref{1int}):
\begin{eqnarray}
\pi_{\aplus} = \frac{1}{2} f(x^+).
\label{pif}
\end{eqnarray}

If we substitute (\ref{aplusrep}) and (\ref{pif}) to (\ref{cancom}),
we obtain the equal-time commutation relations of $f$ and $g$.
Since $f$ and $g$ depend on only $x^+$,
two-dimensional commutation relations 
of $f$ and $g$ are calculated as follows:
\begin{eqnarray}
&& [ f(x^+), f(y^+)] = [ g(x^+), g(y^+)] =0, \nonumber \\
&& [ f(x^+), g(y^+)] = - 2 i \delta( x^+ - y^+).
\label{fandg}
\end{eqnarray}
As seen from $\partial_- f = \partial_- g = 0$,
$f$ and $g$ are the currents which generate the residual gauge
symmetries.
~From (\ref{fandg}) and (\ref{aplusrep}), we can derive the
two-dimensional commutation relation of $\aplus$ as follows:
\begin{eqnarray}
[ \aplus(x), \aplus(y) ] &=& - 2 i
(\Psi^{-1}(f(x^+)))^{\prime}
(x^- - y^-) \delta( x^+ - y^+) \nonumber \\
&=& - 2 i 
(\Psi^{\prime}(\partial_{-} \aplus(x)))^{-1}
(x^- - y^-) \delta( x^+ - y^+) \nonumber \\
&=& - 2 i \left(\frac{\partial^2 \Phi}{\partial (\partial_{-} \aplus)^2}(x)
\right)^{-1}
(x^- - y^-) \delta( x^+ - y^+) \nonumber \\
&=& -  \frac{i}{\pi} 
\left(\frac{\partial^2 \Phi}{\partial (\partial_{-}
\aplus)^2}(x) \right)^{-1} 
\Da(x - y),
\label{acomm}
\end{eqnarray}
where \ ${}^\prime$ is the differentiation of a function and 
$\Da(x)$ is 
defined by 
\begin{eqnarray}
\Da(x) = 2 \pi x^{-} \delta(x^+).
\end{eqnarray}
%
%
%
%
%

It is straightforward to calculate 
multiple commutation relations of $\aplus$ as follows:
\begin{eqnarray}
&& [ \cdots[ [ \aplus(x_1), \aplus(x_2) ], \aplus(x_3) ], 
\cdots, \aplus(x_n)] 
\nonumber \\
&& \quad = (- 2 i)^{n-1} 
\sym_n
 \left[ \Psi^{-1}(z)^{(n-1)} \Biggr|_{z = f(x_i)} \right]
 (x_1^- - x_2^-) \nonumber \\
&& \qquad \times \delta( x_1^+ - x_2^+) \delta( x_2^+ - x_3^+) \cdots
\delta( x_{n-1}^+ - x_n^+), \nonumber \\
&& \quad = \left(- \frac{i}{\pi}\right)^{n-1} 
\sym_n
\left[ \Psi^{-1}(z)^{(n-1)} \Biggr|_{z = f(x_i)} \right]
\nonumber \\
&& \qquad \times \Da( x_1 - x_2) \partial_{-}^{x_2} \Da( x_2 - x_3) \cdots
\partial_{-}^{x_{n-1}} \Da( x_{n-1} - x_n),
\label{multicomm}
\end{eqnarray}
where $\sym_n$ is an arbitrary symmetrization of the arguments $x_1^+,
\cdots, x_n^+$ in $\Psi^{-1}$.
Symmetrization is not necessary in the operator solutions, but it is
necessary when we construct the Wightman functions.
These solutions (\ref{multicomm}) also can be obtained by solving the 
operator Cauchy problems with respect to the multiple commutation
relations.

We can prove that it is sufficient to obtain the above commutation
relations in order to construct all the commutation relations of the 
fundamental fields\cite{AbeNak}.
(\ref{multicomm}) is the exact operator solution of the
quantum nonlinear abelian gauge theory in the Heisenberg picture.

\vskip50pt
\section{The Wightman Functions}
\noindent
Next, we construct the Wightman functions in this theory.
We set the vacuum expectation values of $f$ and $g$ as
\begin{eqnarray}
\wight{f(x^+)} &=& \fvev(x^+), \nonumber \\
\wight{g(x^+)} &=& \gvev(x^+),
\end{eqnarray}
where $\fvev$ and $\gvev$ are arbitrary real c-number functions.
If these one-point functions are non-vanishing, 
Lorentz invariance is broken, but we dare to include nonzero
expectation values 
to consider general situations.
%
%
~From
\begin{eqnarray}
[ f(x), f(y) ] = 0,
\label{fcomf}
\end{eqnarray}
and the energy positivity requirement
we can trivially calculate truncated $n$-point Wightman functions of
$f$. 
For example, the non-truncated two-point Wightman function of $f(x)$
is given by the product of two one-point functions:
\begin{eqnarray}
\wight{f(x)f(y)} = \wight{f(x)}\wight{f(y)} = \fvev(x) \fvev(y).
\label{ffvev}
\end{eqnarray}
In order to obtain the one-point function of $\aplus$, 
we take the following generalized normal-product rule\cite{AbeNak}.

\begin{quote}
The $n$-point Wightman function $W(x_1, \cdots, x_n)$ with 
$x_i{}^\mu = x_{i+1}{}^\mu = \cdots = x_j{}^\mu (i<j)$ 
is defined from 
$W(x_1, \cdots, x_n)$ by setting
$x_i{}^\mu = x_{i+1}{}^\mu = \cdots = x_j{}^\mu$ 
and by deleting the resulting divergent terms in such a way that it be 
independent of the ordering of $i, i+1, \ldots, j$.
\end{quote}
Since the truncated $n$-point Wightman functions of $f$ is trivial,
we obtain the one-point function of $\aplus$ from the above rule as
follows: 
\begin{eqnarray}
\wight{\aplus(x)} &=& \Psi^{-1}(\fvev(x)) x^{-}+ \gvev(x),
\label{onepoint}
\end{eqnarray}
%
%
%
%
%
%
%
Hence (\ref{acomm}) and the energy positivity
requirement 
lead the
two-point Wightman function of $\aplus$ to 
\begin{eqnarray}
\trun{\aplus(x_1) \aplus(x_2)} & = & 
- \frac{1}{\pi} 
\left\langle
\sym_2 \left[
{\left(\frac{\partial^2 \Phi}{\partial(\partial_{-}
\aplus(x_i))^2} \right)^{-1} } 
\right]
\right\rangle
\Dplusa(x_1 - x_2) \nonumber \\
& = & 
- \frac{1}{\pi} 
\sym_2 [(\Psi^{-1}(\fvev(x_i))^{\prime}] \Dplusa(x_1 - x_2),
\label{2point}
\end{eqnarray}
where
\begin{eqnarray}
\Dplusa(x) = \frac{x^-}{x^+- i 0}.
\label{dplus}
\end{eqnarray}

There are plural symmetrization methods of the arguments 
$x_1^+, \cdots, x_n^+$ in 
$\sym_2 [(\Psi^{-1}(\fvev(x_i))^{\prime}]$.
For example, 
\begin{eqnarray}
\sym_2 [(\Psi^{-1}(\fvev(x_i))^{\prime}] 
= \frac{1}{2}[(\Psi^{-1}(\fvev(x_1))^{\prime} 
+ (\Psi^{-1}(\fvev(x_2))^{\prime}], 
\end{eqnarray}
or
\begin{eqnarray}
\sym_2 [(\Psi^{-1}(\fvev(x_i))^{\prime}] 
= \sqrt{(\Psi^{-1}(\fvev(x_1))^{\prime} 
(\Psi^{-1}(\fvev(x_2))^{\prime}},
\end{eqnarray}
are some symmetrizations.
We find any symmetrized solution in (\ref{2point}) is 
consistent with (\ref{multicomm}).
Therefore there are plural Wightman functions
consistent with (\ref{acomm}) and the energy positivity requirement.
In order to determine them uniquely, we need other assumptions.

~From (\ref{multicomm}) and the energy positivity requirement, we can
calculate the $n$-point Wightman functions of 
$\aplus$,
\begin{eqnarray}
&&\trun{\aplus(x_1) \aplus(x_2) \cdots \aplus(x_n)}
= \frac{1}{n} 
\sum^{n!}_{P(i_1, \cdots, i_n)}\sym_n \left[ 
\left( \frac{d}{d Z} \right)^{n-1} \Psi^{-1}(Z) 
\Biggr|_{Z = \fvev(x_i)} \right]
\nonumber \\
&& 
\qquad \qquad 
\times 
\left(- \frac{1}{\pi} \right)^{n-1} 
\Dplusina( x_{i_1} - x_{i_2}) 
\partial_{-}^{x_{i_2}} \Dplusina( x_{i_2} - x_{i_3}) \cdots
\partial_{-}^{x_{i_{n-1}}} \Dplusina( x_{i_{n-1}} - x_{i_n}),
\nonumber \\
\label{nwight}
\end{eqnarray}
%
%
where
\begin{eqnarray}
\Dplusina(x_i - x_j) = \left\{\begin{array}{ll}
\Dplusa(x_i- x_j), & \mbox{if $i<j$} \\
\Dplusa(x_j- x_i), & \mbox{if $i>j$} 
\end{array}
\right.
\end{eqnarray}
and $P(i_1, \cdots, i_n)$ is a permutation of $(1, \cdots, n)$.
The symmetrization $\sym_n$ is arbitrary 
same as (\ref{2point}).
Therefore there are plural solutions consistent with (\ref{multicomm}).

Here, we give some remark on the expression for the $n$-point functions. 
In (\ref{multicomm}), 
we can identically replace the arguments of delta functions.
So there is apparent ambiguity 
at the representations of multiple commutation relations by $D(x)$.
For example, we can have another expression for them as follows:
\begin{eqnarray}
&& [ \cdots[ [ \aplus(x_1), \aplus(x_2) ], \aplus(x_3) ], 
\cdots, \aplus(x_n)] 
\nonumber \\
&& \quad = (- 2 i)^{n-1} 
\sym_n
\left[ \Psi^{-1}(z)^{(n-1)} \Biggr|_{z = f(x_i)} \right]
 (x_1^- - x_2^-) \nonumber \\
&& \qquad \times \delta( x_1^+ - x_2^+) \delta( x_1^+ - x_3^+) \cdots
\delta( x_1^+ - x_n^+), \nonumber \\
&& \quad = \left(- \frac{i}{\pi}\right)^{n-1} 
\sym_n
\left[ \Psi^{-1}(z)^{(n-1)} \Biggr|_{z = f(x_i)} \right]
\nonumber \\
&& \qquad \times \Da( x_1 - x_2) \partial_{-}^{x_1} \Da( x_1 - x_3) \cdots
\partial_{-}^{x_1} \Da( x_1 - x_n).
\label{multicomm2}
\end{eqnarray}
Off course, 
(\ref{multicomm2})
is equal to (\ref{multicomm}).
~From (\ref{multicomm2}), we can derive the different expression for 
of $n$-point Wightman functions:
\begin{eqnarray}
&&\trun{\aplus(x_1) \aplus(x_2) \cdots \aplus(x_n)}
= \frac{1}{(n-2)!n} 
\sum^{n!}_{P(i_1, \cdots, i_n)} \sym_n \left[ 
\left( \frac{d}{d Z} \right)^{n-1} \Psi^{-1}(Z) 
\Biggr|_{Z = \fvev(x_i)} \right]
\nonumber \\
&& 
\qquad \qquad 
\times 
\left(- \frac{1}{\pi} \right)^{n-1} 
\Dplusina( x_{i_1} - x_{i_2}) 
\partial_{-}^{x_{i_1}} \Dplusina( x_{i_1} - x_{i_3}) \cdots
\partial_{-}^{x_{i_1}} \Dplusina( x_{i_1} - x_{i_n}),
\nonumber \\
\label{nwight2}
\end{eqnarray}
Since energy positivity conditions may give us sufficient constraints to
determine the $n$-point Wightman functions uniquely from $n$-ple
commutation relations except for symmetrization ambiguity for the
arguments $x_1^+, \cdots, x_n^+$ in $\Psi^{-1}$,
(\ref{nwight}) and (\ref{nwight2}) should be equivalent.
Indeed we can confirm (\ref{nwight}) and (\ref{nwight2}) are identical 
by using the explicit expression for $\Dplusa(x)$ given by (\ref{dplus}).

\vskip20pt

The exact Wightman functions may break the equations of motion or the
Ward-Takahashi identities in some 
theories\cite{AN3}.
This 'anomaly' arises from regularization of 
divergences of Wightman functions at the same spacetime points.
We have defined the rule to construct the Wightman functions with the
same spacetime points at the sentence after (\ref{ffvev}).
If we define the Wightman functions for the composite fields according
to it, we find that we can 
subtract divergences so as to be consistent with the
equations of motion and the Ward-Takahashi identities in this model as 
follows.

We find that if a truncated Wightman function includes 
$\partial_{-}{}^{x_k} \aplus(x_k)$, 
it does not depend on $x_k{}^-$ from (\ref{nwight}).
And since 
$\partial_{-} \aplus$ commute mutually in two dimension and
the truncated $n$-point functions of $\partial_{-} \aplus$ are zero,
any Wightman function which includes a product of $\partial_{-} \aplus$'s
at the same spacetime point is
non-singular. 
Thus
\begin{eqnarray}
\left\langle \Psi(\partial_{-} \aplus(x_1))
A_+(x_2)\cdots A_+(x_n)\right\rangle, 
\end{eqnarray}
is obtained consistently, and does not depend on $x_1{}^-$.
Therefore 
\begin{eqnarray}
\left\langle \partial_-{}^{x_1} \Psi(\partial_{-} \aplus(x_1))
A_+(x_2)\cdots A_+(x_n)\right\rangle =0, 
\end{eqnarray}
and we can confirm the Wightman functions are consistent with the
equation of motion. 

%
%

\section{The Born-Infeld Theory}
\noindent
The action (\ref{Lag}) reduces to the Born-Infeld theory\cite{BI} 
in two dimension when we set 
\begin{eqnarray}
\Phi(F^{\mu\nu} F_{\mu\nu}) 
= \frac{1}{2 \kappa} \left[ - \sqrt{1+ \kappa F_{\mu\nu}
F^{\mu\nu}} + 1 \right],
\label{phibi}
\end{eqnarray}
where $\kappa$ is a coupling constant.
Then $\Psi$ is written
\begin{eqnarray}
\Psi(\partial_{-} \aplus) =  
\frac{\partial_{-} \aplus}
{\sqrt{1 - 2 \kappa ( \partial_{-} \aplus )^2}},
\label{psibi}
\end{eqnarray}
and $\Psi^{-1}$ is calculated as
\begin{eqnarray}
\Psi^{-1}(f(x^+)) = \frac{f(x^+)} {\sqrt{1 + 2 \kappa f(x^+)^2}}, 
\end{eqnarray}

Expanding the Lagrangian density in power of $\kappa$, 
we obtain the Maxwell theory
at zeroth order:
\begin{eqnarray}
{\cal L} = - \frac{1}{4} F_{\mu\nu} F^{\mu\nu} + \cdots.
\end{eqnarray}
However the Lagrangian density becomes the infinite series of the
fundamental fields.

We can calculate a solution by assuming
\begin{eqnarray}
{\sqrt{1 - 2 \kappa ( \partial_{-} \aplus )^2}} \neq 0,
\end{eqnarray}
if we substitute (\ref{phibi}) and (\ref{psibi}) into the results in
the section 2.

~From (\ref{acomm}), we can derive the
two-dimensional commutation relation of $\aplus$ as follows:
\begin{eqnarray}
[ \aplus(x), \aplus(y) ] &=& - 2 i
\frac{1}{(1+2 \kappa f(x)^2)^{\frac{3}{2}}}
(x^- - y^-) \delta( x^+ - y^+) \nonumber \\
&=& - 2 i [1 - 2 \kappa (\partial_{-} \aplus(x))^2 ]^{\frac{3}{2}}
(x^- - y^-) \delta( x^+ - y^+) \nonumber \\
&=& - \frac{i}{\pi} \sym_{(x,y)} \left[ 
[1 - 2 \kappa (\partial_{-} \aplus(x))^2 ]^{\frac{3}{2}}
\right]
\Da(x - y).
\label{biacomm}
\end{eqnarray}
If $\kappa \rightarrow 0$, the above expression goes to results of 
the Maxwell theory as it should be.
%
%
We derive multiple commutation relations of $\aplus$
from (\ref{multicomm}):
\begin{eqnarray}
&& [ \cdots, [ \aplus(x_1), \aplus(x_2) ], \cdots, \aplus(x_n)] 
\nonumber \\
&& \quad = (- 2 i)^{n-1} 
\sym_n
\left[ \left( \frac{d}{d z} \right)^{n-1} \left( \frac{z}{\sqrt{1 
+ 2 \kappa z^2}} \right) \Biggr|_{z = f(x_i)} \right] 
 (x_1^- - x_2^-) \nonumber \\
&& \qquad \times \delta( x_1^+ - x_2^+) \delta( x_2^+ - x_3^+) \cdots
\delta( x_{n-1}^+ - x_n^+), \nonumber \\
&& \quad = \left(- \frac{i}{\pi}\right)^{n-1} 
\sym_n
\left[ \left( \frac{d}{d z} \right)^{n-1} \left( \frac{z}{\sqrt{1 
+ 2 \kappa z^2}} \right)  \Biggr|_{z = f(x_i)} \right]
\nonumber \\
&& \qquad \times \Da( x_1 - x_2) \partial_{-}^{x_2} \Da( x_2 - x_3) \cdots
\partial_{-}^{x_{n-1}} \Da( x_{n-1} - x_n).
\label{bimulticomm}
\end{eqnarray}

~From (\ref{onepoint}), we
obtain the one-point function of $\aplus$ as follows:
\begin{eqnarray}
\wight{\partial_{-} \aplus(x)} &=& 
\frac{\fvev(x)}{\sqrt{1 + 2 \kappa \fvev(x)^2}}, \nonumber \\
\wight{\aplus(x)} &=& \frac{\fvev(x)}{\sqrt{1 + 2 \kappa \fvev(x)^2}} x^{-}
+ \gvev(x),
\end{eqnarray}
%
%
~From (\ref{nwight}), we can calculate the $n$-point Wightman
functions of 
$\aplus$,
\begin{eqnarray}
&&\trun{\aplus(x_1) \aplus(x_2) \cdots \aplus(x_n)}
= \frac{1}{n}
\sum^{n!}_{P(i_1, \cdots, i_n)} \sym_n \left[ 
\left( \frac{d}{d Z} \right)^{n-1} \left( \frac{Z}{\sqrt{1 
+ 2 \kappa Z^2}} \right) \Biggr|_{Z = \fvev(x_i)} \right]
\nonumber \\
&& 
\qquad \qquad 
\times 
\left(- \frac{1}{\pi} \right)^{n-1} 
\Dplusina( x_{i_1} - x_{i_2}) 
\partial_{-}^{x_{i_2}} \Dplusina( x_{i_2} - x_{i_3}) \cdots
\partial_{-}^{x_{i_{n-1}}} \Dplusina( x_{i_{n-1}} - x_{i_n}).
\nonumber \\
\label{binwight}
\end{eqnarray}
%

\section{Conclusion and Discussion}
\noindent
We have exactly solved the quantum nonlinear abelian gauge
theory 
in two dimension in the light-cone gauge by the canonical-quantization 
operator formalism in the Heisenberg picture.
For the solvability of this theory, we have assumed $\Psi$ be
invertible.
We have calculated the exact multiple commutation relations and 
$n$-point Wightman functions. 
The generalization to the nonlinear non-abelian gauge theory
is trivial.

In this model, two dimensional commutation relation (\ref{acomm}) is 
\begin{eqnarray}
[ \aplus(x), \aplus(y)] \ne 0.
\end{eqnarray}
So we have obtained a new model which 
can be solved by the method proposed by Abe and Nakanishi.

There are symmetrization ambiguities in the Wightman functions
(\ref{2point}) and (\ref{nwight}).
There are plural Wightman functions consistent with multiple
commutation relations and energy positivity.
Therefore the consistency with multiple commutation relations and
energy positivity is insufficient to determine the Wightman functions 
uniquely in this model.
Moreover it is important to clarify the relation between our results
and the analysis in conventional perturbative approach.

In the covariant gauge, we encounter
more general commutation relations than in this paper\cite{II}.
Since it is an example with new commutation relations,
it is interesting to analyze it by 
the operator formalism in the Heisenberg picture.
%
We should develop general mathematical techniques
of the Cauchy problems involving noncommutative quantities\cite{AIN} 
in order to analyze a model with more general commutation relations.

The reguralization procedures in our method is a generalization of
traditional one, for example, the normal ordering of free fields.
And ours are applicable to not only free fields but general Heisenberg
fields.
Therefore it is interesting to apply to the gravitational theory, which
is essentially
non-polynomial and non-renormalizable. 
%
This method for solving quantum theory will be useful to treat
non-polynomial quantum field theories.






\section*{Acknowledgements}
The author thank Prof.M.Abe and Prof.N.Nakanishi 
for discussions and comments about the present work. 
He express gratitude to Prof.M.Abe
for reading the manuscript carefully.

\newcommand{\bibit}{\sl}



\vfill\eject
\end{document}